# How balloons lose their charge by collecting ions from the air

*J. Pantaleone, Dept. of Physics, University of Alaska Anchorage, Anchorage, Alaska 99508*

The air we breathe contains radioactive particles. Each time one of these particles' decays, it creates hundreds of thousands of ionized air molecules. These ions make the air conductive, which is one reason insulated surfaces lose charge over time. Here we observe the discharge of two statically charged balloons by measuring the repulsive force between them over time. The initial rates of charge loss are relatively large, especially at higher humidities. This large rate of charge loss is likely dominated by the emission of water ions from the balloon's surface. Approximately 1 hour after charging, the charge loss rates are much smaller and agree with predictions calculated from the radon concentration in the air and the size of the room, for reasonable aerosol concentrations.

## I. INTRODUCTION

A popular electrostatics activity is to observe the repulsion between two charged balloons. The balloons are hung from strings and rubbed with fur to become negatively charged, and then the geometry of the system is measured. Coulomb's law can be demonstrated this way and, if the weights of the balloons are known, the average charge on the balloons can be calculated.[1-3] Here, this experiment is videotaped with a smartphone to record how the geometry changes with time. From this the decrease of charge with time is found.

There are several processes that can cause charges on insulators to decrease. These include conduction through the bulk of the insulator, conduction along the surface of the insulator, emission of ions from the insulator to the air, and the attraction of oppositely charged ions in the air to the insulator. For the balloon experiment discussed here, the strings are long, very thin, and made of a material with a large resistivity. Thus the conduction through the strings should be negligible,[4] and the focus here is on the interaction of the balloons with the air.

Balloons are made of polymers, and many studies have examined the electrostatic properties of polymers. These studies have found that polymers lose charge at rates that increase with the humidity of the air, and that depend on the type of polymer and the sign of the charge.[5] Because these rates depend on the type of polymer, they can't be due solely to the attraction of ions from the air since that process depends on the electric field far from the surface and should not be sensitive to the local surface properties. Thus these observations are at least partly ascribed to emission of water ions to the air,[5,6] a process relevant for the balloon experiments discussed here. The effects of water vapor in the air on the buildup and decay of charges on dielectrics is an ongoing area of research.[7-9]

Charge loss by attraction of opposite sign charges from the air is a relatively well understood process. This process, measured using electroscopes, was the main method used to quantify the intensity of ionizing radiation in the pioneering work of Wilhelm Roentgen, Henri Becquerel, and Victor Hess.[10] However, this process is rather complicated when applied to ion collection in a room. In that case it depends on the size of the room containing the experiment, the aerosol concentration, and slightly on the humidity of the air. Qualitatively, more ionizing radiation and bigger rooms mean more ions to be collected and so a larger rate of charge loss. However large aerosol concentrations and relative humidities mean the ions in the air tend to be larger, and larger objects have smaller drift velocities, so there is a smaller rate of charge loss in large rooms.[11] In this article, a model is developed to describe ion collection from the air by charged balloons in a room. Inexpensive radon and gamma ray detectors are used to calculate the rate of ion production in the air of the room, and this is used to calculate the expected initial and steady state electrical currents to the charged balloons.

These predictions are compared to the results of experiments, performed in summer and winter, and for a range of balloon sizes. In summer, the relative humidity is higher while the level of ionizing radiation is slightly lower than in winter. The rates of charge loss for the same size charge are observed to initially be three times higher in summer than winter, presumably due to emission of water ions. At one hour after charging the rates are slightly smaller in summer and this suggests the late time behavior is dominated by the collection of ions from the air. This conclusion is reinforced by the observed charge loss rates agreeing with the theoretical predictions for reasonable aerosol concentrations.

The discussion here is of general interest for several reasons. First, balloons are an everyday item and charged balloons are widely used for science activities. More generally, electrostatic experiments are common in introductory physics courses, and the discussion here describes ways the environment can influence these experiments. In addition, the experiment demonstrates why ionizing radiation has that name, and it complements cloud chamber observations of ionizing radiation.

This paper is organized as follows. Section I introduces the topic and outlines the paper. Section II reviews the basics of ions in the air: II.A shows how to calculate the number of ions created by background ionizing radiation, II.B describes typical equilibrium ion concentrations and lifetimes without an electric field, and II.C describes how a charged object collects ions. Section III describes the experiment and presents the results. Section IV compares the experimental results to the theory, and section V contains a few concluding remarks. The supplementary material contains additional details.

## II. THEORY

### A. Creating ions in the air

It takes energy to create an ion in the air. Waterfalls and fires produce ions, but away from these, the main sources of ionizing energy are radioactive decays and cosmic rays. Both of these create energetic particles---alphas, betas, gammas, and muons---that ionize the molecules around them as they move through the air.[12] Radioactive decay of particles in the soil can also produce ions when they emit gamma rays that travel through the soil and into the air. The total rate of ion pair production in the air per unit volume, $c$, can be written as

$$c = c_r + c_{cr} + c_g \quad , \tag{1}$$

where the subscripts denote the different ionizing processes: (r) the charged particles emitted by a radioactive decay in the air, (cr) cosmic rays and (g) gamma rays from radioactive decays nearby.

Cosmic rays hitting the upper atmosphere produce a cascade of secondary particles traveling downwards. Almost all of these particles are stopped by the atmosphere before they reach the Earth's surface. However, muons produced in these cascades are electrically charged particles that can reach the surface of the Earth and produce ions there.[13] At ground level, $c_{cr} \approx 2$ cm$^{-3}$ s$^{-1}$ for the high geomagnetic latitudes where this experiment was performed.[14]

Gamma rays in the air near the surface of the Earth come mostly from radioactive decays of uranium and its progeny in the top 30 cm of the soil and in the air. The intensity of these gamma rays depends on the local geology. The average value of $c_g$ over the land areas of the Earth is 4 cm$^{-3}$ s$^{-1}$ up to 200-300 m from the surface.[14] An inexpensive gamma ray spectrometer measured the intensity in the experimental room to be $4.0\times10^{-6}$ μR/h, for both summer and winter. Since the density of the air in the room was 1.17 kg/m$^3$ (see Section III) this corresponds to $c_g \approx 2$ cm$^{-3}$ s$^{-1}$.

Table 1. Unstable isotopes in the $^{238}U$ decay chain starting with radon, $^{222}Rn$, and ending with $^{210}Pb$. The average energy of the charged particle they emit are calculated from the tables in Ref. 15 and are used in Eq. (2). These decays also emit gamma rays, but that contribution to $c$ is included in $c_g$.

| Isotope | Charged particle emitted | Half-life | Average energy (MeV) |
|---|---|---|---|
| $^{222}Rn$ | alpha | 3.82 d | 5.49 |
| $^{218}Po$ | alpha | 3.05 min | 6.00 |
| $^{214}Pb$ | beta | 26.8 min | 0.22 |
| $^{214}Bi$ | beta | 19.9 min | 0.63 |
| $^{214}Po$ | alpha | 164 μs | 7.69 |
| $^{210}Pb$ | beta | 22.3 years | $6.5\times10^{-3}$ |

The largest contribution to $c$ comes from the radioactive decay of particles in the air. The most abundant radioactive isotopes in the air are usually radon[16] and its progeny, see Table 1. Radon is a noble gas, and so after being created it may escape from the soil into the air. The rate of radon-related ion production, $c_r$, is calculated by adding up the rate of energy deposition in the air from the emitted alphas and betas and dividing that energy by $E_{ion} = 34$ eV, the average energy required to produce an ion pair in air from ionizing radiation:[17]

$$c_r \approx \frac{1}{E_{\text{ion}}} \sum_i C_i E_i g_i \quad . \qquad (2)$$

Here the subscript $i$ denotes the radioactive species that emits the charged particle, with $C_i$ the activity concentration, $E_i$ the average energy of the particle (see Table 1), and $g_i$ a geometric factor describing what fraction of the particle's path is through the air.[18] The factor $g_i$ is 1 if the parent nucleus is in the air when it decays and ½ if it is attached to a flat, macroscopic surface. The decays of $^{210}Pb$ and its progeny are not included in Eq. (2) because $^{210}Pb$ has a much longer half-life. These particles will have become attached to a surface before they decay and are presumably removed through regular cleaning of the room.

To specify the $C_i$ needed for Eq. (2), we assume that the radon in the air is in secular equilibrium with all the species in Table 1 before $^{210}$Pb, so all their activity concentrations are the same. For the present experiment, the activity of radon in the experimental room was measured (see Sec. III) and averaged over the days when experiments were performed (without the times the balloons were charged). These values are given in Table 2. Also shown in Table 2 are the corresponding values for $c$ assuming these daughters of radon are all in the air, or all on a surface. The values are generally much larger than the contribution from cosmic rays and gamma rays, $c_{cr} + c_g = 4$ cm$^{-3}$ s$^{-1}$.

Table 2. Total ion pair production rate, $c$, for the summer and winter radon concentrations observed here, and also for the US national indoor average.[19] Separate values are given for when the progeny of radon are on a surface and in the air.

| | Summer | | Winter | | US average | |
|---|---|---|---|---|---|---|
| $C_{radon}$ (pCi/L) | 0.39 | | 0.74 | | 1.3 | |
| | Surface | Air | Surface | Air | Surface | Air |
| $c$ (cm$^{-3}$ s$^{-1}$) | 9.4 | 12.5 | 14 | 20 | 22 | 32 |

The fraction of radon's progeny attached to a surface will be small when the balloons are first charged, but it grows with time. This is because emission of an alpha particle leaves the daughter atom positively charged at the end of its recoil path 88% of the time.[20-22] Subsequently, this daughter atom tends to remain positively charged because it has a lower ionization energy than the surrounding air molecules.[23] The mobilities and lifetimes of the radioactive and nonradioactive ions are similar, so the radioactive ions will be collected by the balloons, and then tend to remain on the balloons' surface.[24] Observing these radioactive daughters on the surface of a balloon is a popular classroom activity.[25-27] The "Surface" values in Table 2 are approximately applicable in the limit when the charge on the balloons is large enough to collect all of the positively charged ions in the room, and the collection time is long enough that the equilibrium abundance of radon's daughters that remained in the air have decayed away.

## B. Ion properties without electric fields

When an ion is first created in the air, it quickly collects a few molecules around it and becomes a "small" ion, with a diameter less than 1.6 nm.[28] Small ions have large mobilities and are primarily responsible for the electrical conductivity of the air. Small ions become "large" ions when they attach themselves to aerosols, which typically have diameters greater than 7.5 nm,[28] and so much smaller mobilities.

Table 3. Typical equilibrium concentrations ($n^*$ and $N^*$) and lifetimes ($t_S$ and $t_L$) for small and large ions, respectively. These assume an ionization rate of $c$ =20 cm$^{-3}$ s$^{-1}$ (see Table 2), an aerosol radius = 10 nm, and an aerosol concentration of 2x10$^5$ cm$^{-3}$, which are typical for an urban environment.[29] The "neutral fraction" is the fraction of aerosols that are neutral.

| | |
|---|---|
| $n^*$ | 330 cm$^{-3}$ |
| $N^*$ | 14,000 cm$^{-3}$ |
| Neutral fraction | 86% |
| $t_S$ | 16 s |
| $t_L$ | 11 min |

The abundances and lifetimes of small and large ions are determined by a balance between their rates of creation, transformation and destruction by recombination. These microscopic processes can be modeled using the law of mass action.[14] Table 3 gives typical values expected for the environment of the experiment, calculated assuming no electric fields or macroscopic surfaces. See supplementary material for details.

**C. Collection of ions by charged balloons.**

To understand the collection of ions by charged balloons, we start by calculating the time it takes for a positive ion to travel from a corner of the room to the negatively charged balloons hanging at the center of the room. With an electric field, $\vec{E}$, present, a positive ion will diffuse through the air along a field line with a drift velocity, $\vec{v}$, of

$$\vec{v} = \mu\vec{E} \tag{3}$$

where $\mu$ is the mobility. The travel time for an ion is not sensitive to the size of the balloons or the distance between them, as long as the room is large compared to either of those characteristic lengths, since most of the ion's travel time occurs when the ion is far from the balloons and therefore moving slowly. Thus a reasonable approximation is to take the electric field in the room as that from a single point charge of magnitude $Q$, where $Q$ is the sum of the charges on the two balloons. Then the travel time of an ion starting a distance $r$ away from a balloon with characteristic size $R$ is

$$t_{\text{travel}} = -\int_r^R \frac{dr\prime}{v} \sim \frac{-1}{\mu kQ}\int_r^R r'^2 dr' \sim \frac{1}{3}\frac{r^3}{\mu kQ} \tag{4}$$

where Eq. (3) was used and $k$ is coulomb's constant. Evaluating Eq. (4) for the largest possible travel distance for the present experiment gives the travel times in Table 4, $t_{\text{travel,max}}$. Table 4 also contains other quantities relevant to ion collection by electric fields.

Table 4. Typical ion collection scales for the present experiment for the small and large ions. The average mobility, $\mu$, of the small positive ions is used,[30,31] while the mobility of the large ions is calculated from the Stokes equation with Millikan's corrections.[32] The maximum travel times are found using Eq. (4) with $Q \approx 1.4$ μC and $r_{\text{max}} \approx 4.2$ m (see Sec. III). The quantity $Q_{\text{room}}^*$ is the charge in the air of the room of each sign at equilibrium, which is $eN^*V_{\text{room}}$ for the large ions and $en^*V_{\text{room}}$ for the small ions where $V_{\text{room}}$=81 m$^3$ is the volume of the room (see Sec. III). The approximate typical initial current to the balloons from the equilibrium ions is given by Eq. (5).

| | Small | Large |
|---|---|---|
| $\mu$ | 1.3 cm$^2$ V$^{-1}$ s$^{-1}$ | 5.5x10$^{-3}$ cm$^2$ V$^{-1}$ s$^{-1}$ |
| $t_{\text{travel,max}}$ | 15 s | 1.0 h |
| $Q_{\text{room}}^*$ | 4.3x10$^{-3}$ μC | 0.18 μC |
| $I_{\text{initial}}$ | N.A. | 0.20 nA |

The typical maximum travel time, $t_{\text{travel,max}}$, for the small ions is roughly the same as the typical small ion lifetime in Table 3, and much shorter than the 2 or 3 minutes it takes to charge the two balloons by rubbing and to start video recording. Thus, approximately *all* of the equilibrium small ions initially in the room will be removed by the balloons *before* the current to the balloons is measured! This means that the initial current of small ions will not be observable, and so it is not given in Table 4. This current would not be observable, in any case, because the initial amount of charge on the small ions in the room, $Q_{\mathrm{room}}^*$, is much smaller than the charge on the balloons.

For the large ions the situation is different. The large ions move much slower and have much longer lifetimes. To estimate the initial current from the large ions, it is reasonable to approximate the large ion concentration as a constant over the first few minutes of observations. For a constant density of charged particles, the current of large ions towards the balloons is found by summing the flow of positive ions over a closed surface around the balloons

$$I_{\mathrm{initial}} \approx eN^* \oint \vec{v} \cdot d\vec{A} \approx eN^* \mu \, 4\pi k Q \qquad (5)$$

where $e$ is the unit of charge. In the last step, Eq. (4) was used and the surface integral was done using Gauss's law. Equation (5) is known as the Gunn relation,[33] and the typical value for this current is given in Table 4 for the large ions. This current will continue until the large ion concentration is depleted due to collection on the balloons and to large ion recombination. This process will take several minutes, after which the large ion current will remain very small because large ions are only created from small ions and there are very few of the small ions left in the room.

After the abundances of ions are depleted around the balloons, there will still be a small current of small ions to the balloons because small ions are constantly being created by ionizing radiation. Experiments[34,35] show that this current is well described by

$$\frac{dQ}{dt} \approx -ecV \quad , \qquad (6)$$

where $c$ is the ion pair production rate, as described in Sec. II.A, and $V$ is the volume inside of which $t_{\mathrm{travel}} < T$, where $T$ is a constant of order the small ion lifetime, $t_S$. Equation (6) says that the ions near the balloons are collected as fast as they are produced by radioactivity, while outside of this volume the effects of the electric field are weak, so ions are dominantly removed by microscopic processes, and the current from that region is negligible, see Fig. 1

To illustrate how the volume $V$ depends on charge, we can calculate it for the geometry of a rectangular room around a point charge $Q$. The room is taken to have vertical height $H$ and equal widths $W$, with $W$ substantially greater than $H$. Inverting Eq. (4) and using $T$ for the limiting travel time gives

$$r_{\mathrm{collection}} = [3\mu k Q T]^{1/3} \qquad (7)$$

for the distance from the center over which ions are collected. Then, as $Q$ increases $r_{\mathrm{collection}}$ increases, and the volume around the point charge where the field is strong is approximately

$$V \approx \begin{cases} 4\pi r_{\mathrm{collection}}^3/3 & r_{\mathrm{collection}} < H/2 \\ H\pi r_{\mathrm{collection}}^2 & \frac{H}{2} << r_{\mathrm{collection}} < W/2 \\ V_{\mathrm{room}} & W/\sqrt{2} < r_{\mathrm{collection}} \end{cases} \quad . \qquad (8)$$

For small charges the volume is approximately spherical, for intermediate charges it is approximately cylindrical, and at large charges it is the entire volume of the room. While $T$ is not known a priori, it will be chosen to fit the data.[34,35]

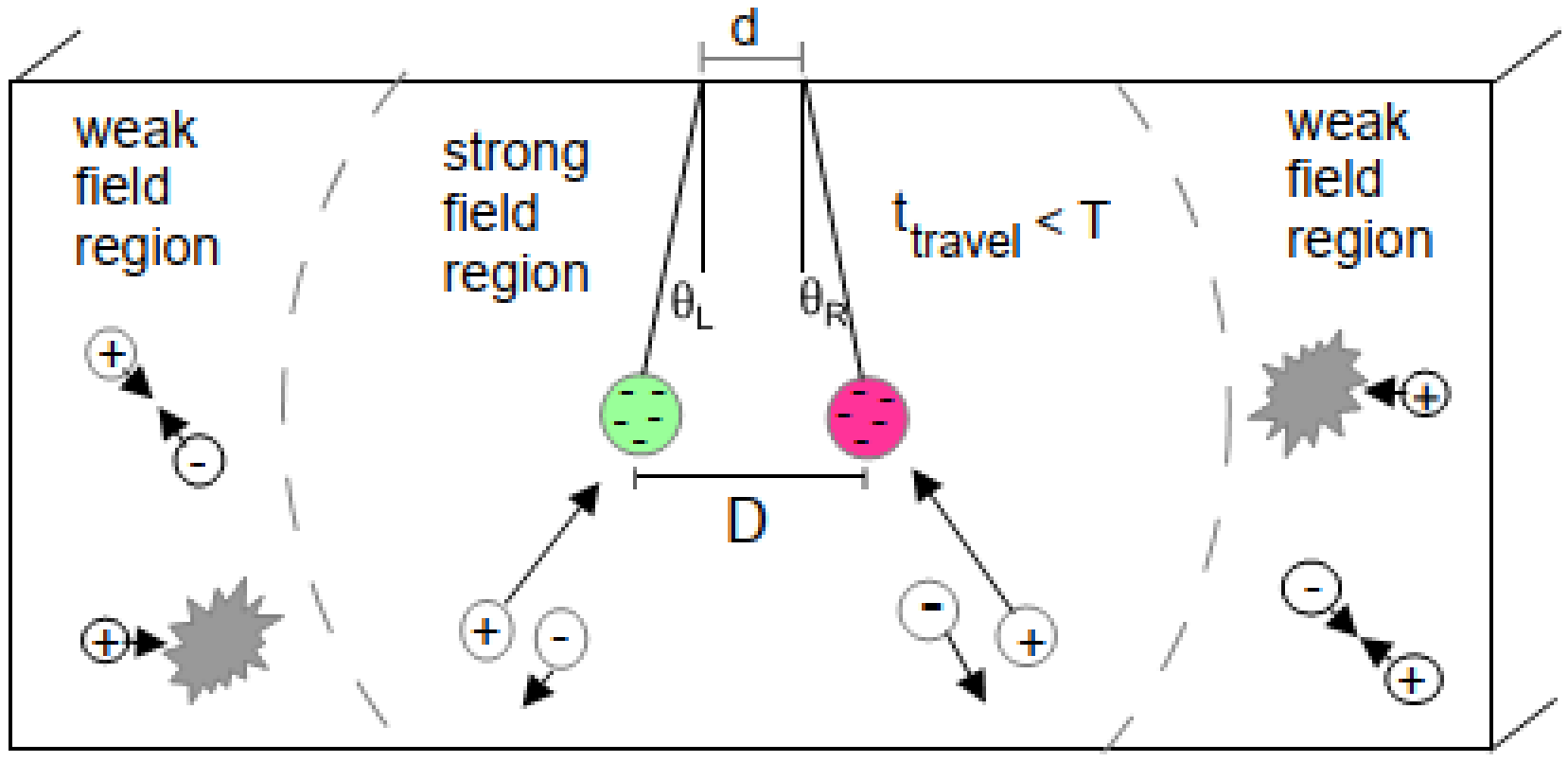

Fig. 1. Cross-section of room containing balloons. Inside the dashed curves is the volume $V$ where small ion travel to the balloons is faster than small ion removal by collisions.

For comparison with the data measured here, a slightly more sophisticated method was used to find the charge dependence of $V$. The room volume was discretized to a 3-dimensional grid of points and the travel time calculated numerically for each point. These travel times were then sorted and summed to find the number of points as a function of the travel time. This requires only a scaling to give the volume as a function of $r_{\text{collection}}$. This numerical method was used to account for several small effects. In order of decreasing importance, these small effects are: that the room's base is not exactly square, that the balloons are slightly off center (to maximize the distance from them to the camera tripod and other structures in the room), that there are two balloons with a small separation between them, and that there is a small image charge of the balloons in the concrete floor with soil under it. Distortion of the electric fields from the dielectrics making up the ceiling and walls are neglected, because their thicknesses are small compared to their distance to the balloons (see supplementary material for more details).[36]

## III. EXPERIMENTAL METHODS AND RESULTS

The experiments were performed in an empty garage in the basement of a house. The room had a maximum height of 2.8 m, and maximum horizontal dimensions of approximately 5.9 m by 6.0 m. The approximately square floor plan was reduced by various structural features that were not easily removed: one corner of the room was cut-off by a diagonal wall, a wooden workbench was along another wall, two small HVAC systems were on two walls, and the ceiling had 2 air ducts and two wooden beams. The total air volume in the experimental room was 81 m$^3$. The floor of the room was concrete with soil underneath, the ceiling was drywall, and the walls were a mixture of plastic sheeting, drywall, concrete block and garage door. The plastic sheeting was used to isolate the air in the room from the rest of the house.

The environmental conditions for the experiment are given in Table 5. The gamma ray intensity was measured with a Radiacode 103 detector. The activity of radon in the room was measured with an EcoQube radon detector.[37,38] All winter measurements were done during the month of January, in a subarctic climate, with exterior temperatures well below freezing and the ground

frozen---conditions that increase the radon activity in heated buildings. Summer measurements were made during the last week of June and first week of July, when exterior temperatures are typically near the highest.

Table 5. The measured environmental conditions in the experimental room, averaged over the days that balloon charge measurements were made.

| | Winter | Summer |
|---|---|---|
| Temperature | 17 ± 1 °C | 19.4 ± 0.4 °C |
| Relative humidity | 28 ± 2% | 52 ± 1% |
| Atmospheric pressure | 979 ± 8 mb | 981 ± 2 mb |
| Gamma ray exposure | 4.0 μR/h | 4.0 μR/h |
| Radon concentration | 0.74 ± 0.06 pCi/L | 0.39 ± 0.04 pCi/L |

To measure the charge on the balloons, the set-up shown in Fig. 1 was used. The balloons were first inflated with an air compressor, then the mass of each in air, $m_j$, was measured with a lab balance, where the subscript $j$ denotes L (left) or R (right). The balloons were then tied to polyethylene fishing line (diameter 0.66 mm) and suspended half-way between the ceiling and floor.

After rubbing the balloons with a rabbit fur for about a minute each, the balloon positions were recorded with a smart phone camera at a low frame rate from a distance 2.5 m away for 60 minutes. The interval between rubbing ending and video recording beginning took no more than a minute, just enough time to allow fluctuations in the balloons' positions to become unnoticeable. The experimenter exited the room with the rabbit fur immediately after video recording started.

The video was analyzed using the software Tracker.[39] A plumb line was used to determine the vertical direction, and markings on it were used to calibrate distances. Frames 40 seconds apart were examined to find the angle of each string from vertical, $\theta_j$, and the distance between the centers of the balloons, $D$. The geometric average of the balloon charges in each frame was found using the observations for each balloon and the equation

$$\sqrt{q_R q_L} = D\sqrt{\frac{m_j\, g\, tan\theta_j}{k}} \qquad (9)$$

where $q_j$ is the charge of the balloon on side $j$, $g$ is the acceleration due to gravity and $k$ is Coulomb's constant. This equation follows from Coulomb's law, Newton's third law, and the conditions for static equilibrium. The values for $\sqrt{q_R q_L}$ from the observations of the left and right balloons should be equal, and the difference between them was used to monitor the error in the measurements. The total central charge, $Q=q_L+q_R$, was approximated by $2\sqrt{q_R q_L}$ with a relative uncertainty that started around a few percent and decreased over time (see supplement for details). The distance between the strings at the top, $d$ in Fig. 1, was chosen to be about 35 cm in order to make the uncertainty in $Q$ from the estimated uncertainties in $D$ and $\theta$ comparable. The uncertainty in $Q$ due to image charge effects[36] from the concrete floor and dirt underneath it is estimated to be less than 2%, so no correction for this effect was applied (see supplementary material for details).

Three different sizes of balloons were used with package labels reading 7 inch (17.8 cm), 9 inch (22.8 cm), and 12 inch (30.4 cm). These balloons were all from the same manufacturer. The actual size of the balloons varied with the amount of inflation, and two balloons of similar size were always used in a charge measurement. For each measurement, new balloons were always used.

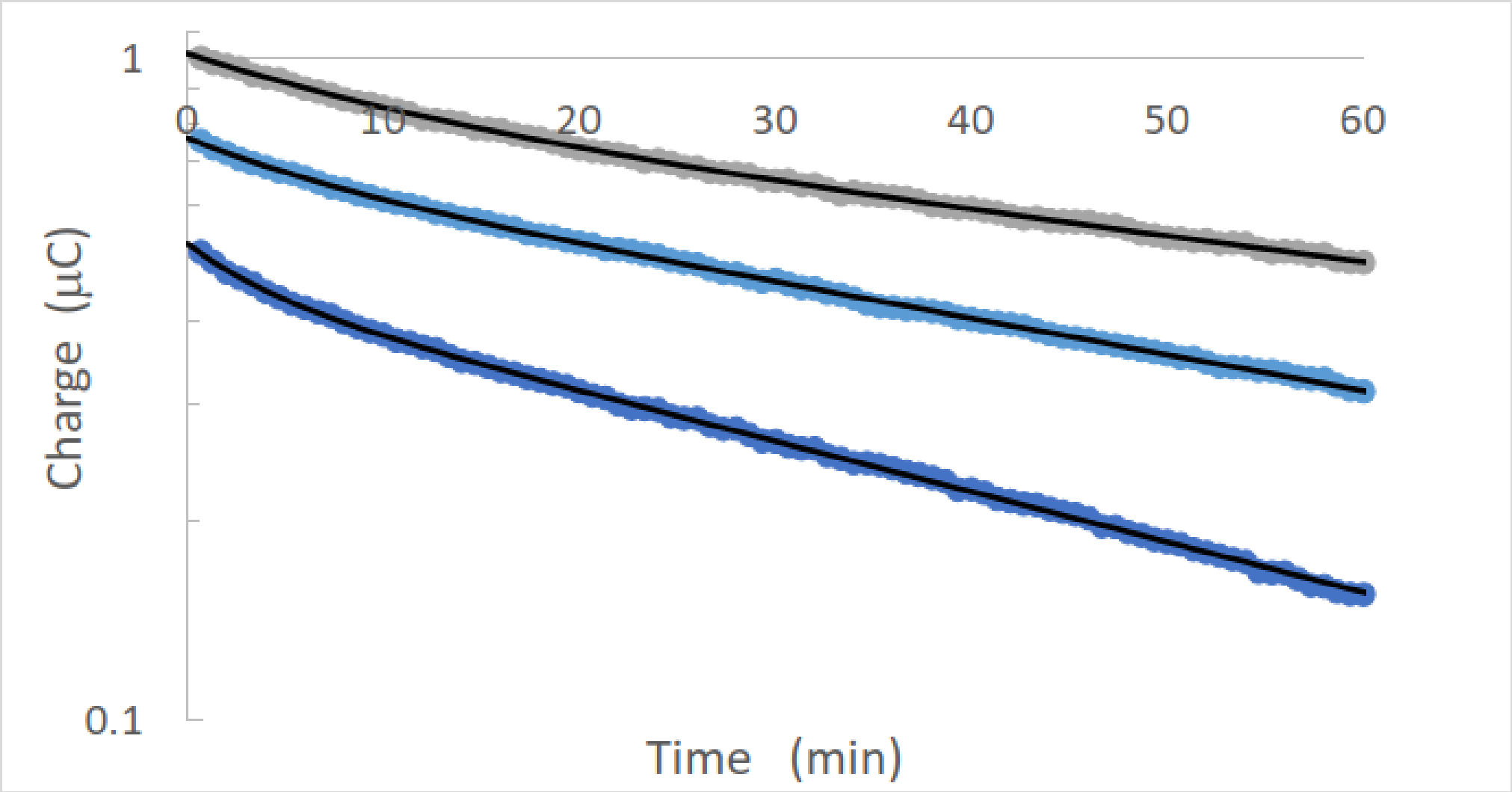

Fig. 2. Log-linear plot of charge on a balloon versus time for a large, medium and small balloon (top to bottom), observed in the winter. The thin, solid black curves are fits to the data.

Figure 2 shows three typical observations of charge versus time. In general, the rate of charge loss is relatively fast for the first few minutes, followed by a slower decay. The time dependence of the data can be equally well fit by several different types of functions, so it is not possible to draw reliable conclusions about the mechanisms responsible for the current to the balloons from a single observation of charge versus time.

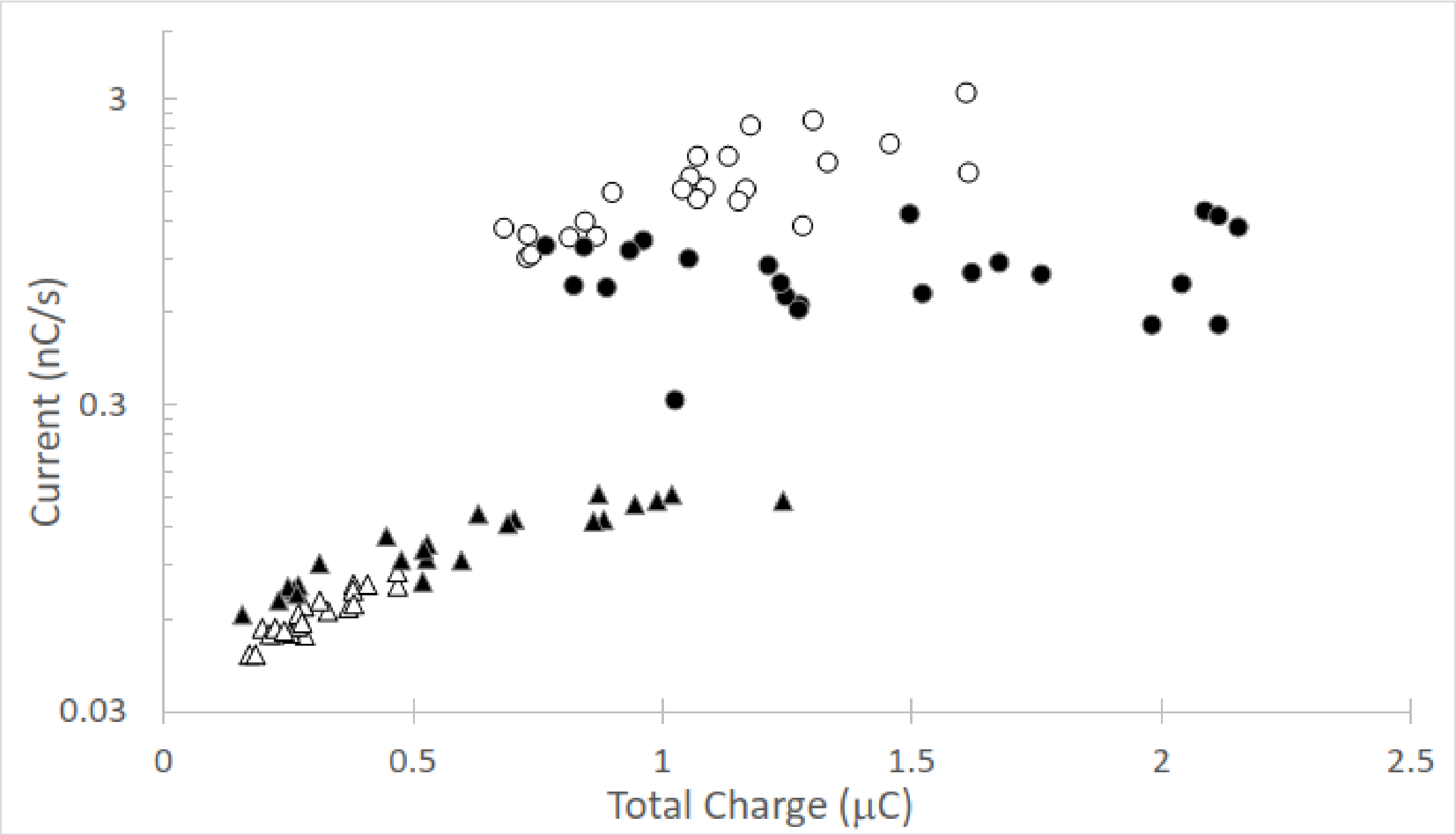

Fig. 3. Log-linear plot of total current towards the two balloons versus the total charge on the two balloons. Open symbols correspond to summer while filled symbols to winter, circles and triangles correspond to measurements at $t$ = 0, and 60 minutes, respectively.

The observations are plotted as current versus charge in Fig. 3. This is a useful way to plot the data because the electric field is proportional to the charge, and it is the electric field that causes the current. The current to a balloon is the slope of the charge versus time data in Fig. 2, and a mathematical method was used to calculate a smooth value for the slope. The charge versus time data was fit to the sum of two exponentials (two half-lives and two amplitudes, see supplementary for additional details). This method yielded fits with the fraction of variance unexplained = $(1-R^2) < 1\text{x}10^{-3}$. It was these fits that were used to construct Fig. 3.

## IV. COMPARISON OF OBSERVATIONS TO THEORY

Figure 3 shows the charges and currents at $t$=0 minutes and 60 minutes after charging, for summer and winter. The initial charges are larger for winter, indicating a larger charge density on the balloon's surface created by rubbing.[40] The initial currents (circles) for both summer and winter are larger than the 0.20 nA expected at $Q$=1.4 μC from large ions (see Table 4) with the summer being the largest. Thus these observations disagree with the large ion collection model developed here, especially since in the summer environment the aerosol concentration is expected to be smaller (so smaller large-ions currents) and the humidity is higher (so slightly smaller ion mobilities). This suggests that another mechanism is at work initially, such as emission of water ions from the surface of the balloons.[5,6] The fluctuations in the initial currents might be due to different amounts of adsorbed water on the surface of the balloons, or variations in the time it takes to start video recording.

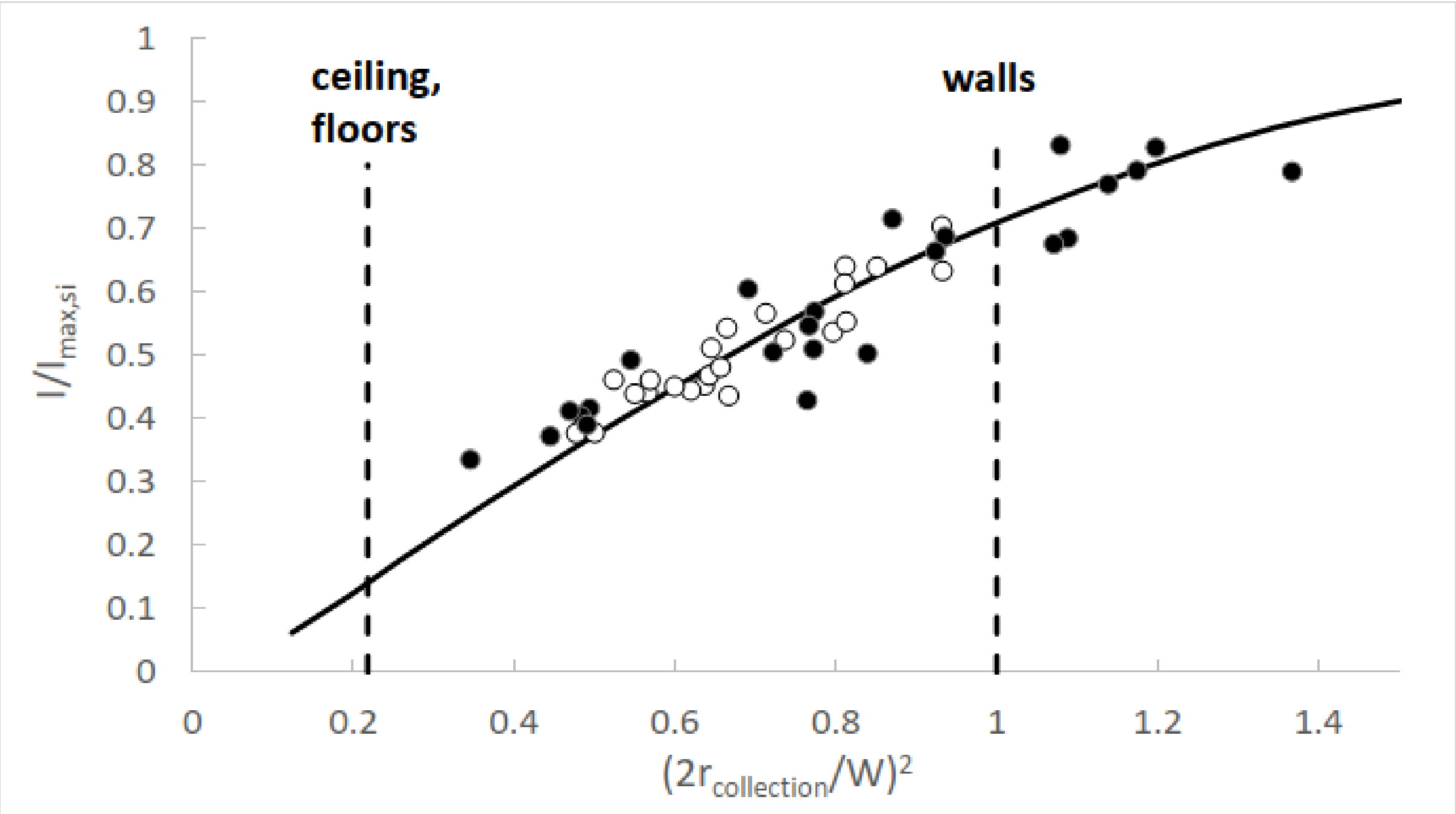

Fig. 4. Plot of data at t=60 min with open (filled) circles observations for summer (winter). The vertical axis is the measured current divided by the maximum possible current from small ions, $I_{max,si} = ecV_{room}$, which is calculated using the "surface" values of $c$ in Table 2. The horizontal axis is the ion collection distance, $r_{collection}$, divided by the average distance from the balloons to the walls, $W/2$, quantity squared. This ratio is $([3\mu kQT]^{1/3}/\ 3\ \text{m})^2$ and is calculated using seasonal values for the mobilities[30] and the values for $T$ of 17 s for summer and 9 s (chosen to fit the data to the curve). The solid curve is the ion collection volume divided by the room volume, $V/V_{room}$, and the vertical dashed lines indicate the transition regions in this curve from surfaces in the room (see Eq. (8)).

In contrast to the $t$=0 data, the $t$=60 minute currents in Fig. 3 (triangles) are much smaller, have smaller relative fluctuations, and are more consistent between summer and winter. To see if these data are consistent with the small ion collection model described in Section II, the data is replotted in Fig. 4 with the axes modified. The vertical axis of $I/I_{max,si}$ is, according to the small ion collection model (see Eq. (6)), the fraction of the room volume from which small ions are collected, $V/V_{room}$, and should always be less than 1, in agreement with the data. Along the horizontal axis, the data lies in the range where the collection volume is always cut-off by the ceiling and floor, and also slightly by the walls for the largest few charges. Over this range the collection volume is approximately given by Eq. (8), intermediate case

$$\frac{V}{V_{\text{room}}} \approx \left(\frac{H\pi r_{\text{collection}}^2}{HW^2}\right) = \frac{\pi}{4}\left(\frac{2r_{\text{collection}}}{W}\right)^2$$

to within 10%.

The seasonal maximum travel times, $T$, used to scale the data in the horizontal direction, were chosen to best fit the data to the curve. The $T$ values are comparable to the typical small ion lifetime given in Table 3, $t_S$ =16 s. The summer $T$ value is larger than the winter value, in agreement with expectations since in summer there are fewer aerosols due to far less burning of fossil fuels to heat homes in the subarctic climate. The data fit the curve very well, and the small fluctuations in the measurements about the curve may be due to fluctuations in the daily radon or aerosol concentrations due to the weather.

Figure 4 shows that for the larger charges the scaled currents are close to the limiting value of 1. Thus these balloons are collecting almost all of the positive small ions created in the room at $t$ = 60 minutes, and so also at all the times before this.

## V. CONCLUDING REMARKS

Charged balloons collect small ions from the air as fast as they are produced by ionizing radiation throughout a volume around them. This volume is limited by surfaces in the room and by the distance a small ion can travel in a lifetime. The observed currents at 60 minutes after charging are well described by this model, with the assumption that other ion sources are negligible. This model should be tested, and there are many possible student experiments that can do this since the most expensive piece of equipment used here was a smartphone, to film the balloons. An EcoQube radon detector and a Radiacode gamma ray detector were also used, but these items are now readily available online at costs much less than a smartphone, and have many uses in introductory physics.[27]

One possible project would be to study how the aerosol density affects the rates of charge loss. Aerosols could be added to the air (e.g. with a can of smoke/fog) and that would decrease the distance the small ions can travel and so should decrease the balloons' long-term rate of charge loss. Conversely, aerosols could be removed from the air with filters, and this should increase the long-term rate of charge loss---if the filters did not also remove radioactive isotopes from the air.

The aerosol concentration was not independently measured here because, although commercial aerosol sensors are readily available to measure the particulate *mass* density, and they indicated higher concentrations in the winter for the environment of this experiment, the present experiment depends on the *number* density of aerosols and their size distribution.

Another possible project would be to study the effects of changing the room volume. With a somewhat smaller room than used here, $r_{\text{collection}}$ would be larger than the distance to the room perimeter at 60 minutes. Then the measurements would be independent of the aerosol density. This small volume limit is the condition under which electroscopes are traditionally used as radiation detectors. However, with smaller volumes, the ion attachment to the walls becomes a more significant source of ion removal.[18] For the model used here, the effect of the walls was "absorbed" into the model's fit parameter, $T$, but its importance could be studied by varying the room volume in which the ions are collected.

An uncertainty in the model concerns what fraction of the radioactive ions in the room are on the surface of the balloon. In Fig. 4 it is assumed that all radon daughters are on the balloons' surfaces at $t$=60 min. If instead it is assumed that the daughters are all in the air, then the ion production rates are about 25% higher, and the model can still fit the observations with smaller values of $T$ (see supplementary material). A more accurate value for the fraction of radon's daughters in the room that are on the balloon is desirable, but this is difficult to determine. The freshly produced radioactive ions are expected to have a mobility and a lifetime similar to those of the small ions, which is confirmed by many measurements. However the uncertainty in the value of their mobility is large, possibly due to a sensitivity to trace gases in the air.[23,41,42] This makes more precise predictions difficult.

Another possible project would be to study the "Radioactive Balloon," a popular classroom activity. This involves collecting ions from the air with a charged balloon and then observing the balloon's radioactivity with a Geiger counter[25-27] or gamma ray detector. The radioactivity of the balloon should be calculable using the ion collection model described here. Also, measuring how this radioactivity varied with the balloon's size, or with the volume of the room, would probe the fraction of radioactive ions that are charged and their mobility.

It would also be interesting to study how the charge measurements depend on the properties of the balloons. For example, inflating the balloons with a lifting gas would increase their sensitivity to charge.[1,2] Also, using balloons made of a material besides latex (such as vinyl, mylar or synthetic rubber) could change the amount of water adsorbed on the surface and so change the initial current values but not the long-term values. A similar effect might be achievable for a latex balloon by a surface treatment with a water absorbing material such as corn starch, or with a silicone based spray.

Another possibility for student projects would be to study the conductivity of the air. The present experiment only involves the flow of positive ions, so it does not measure the conductivity of the air, since that depends on the densities and mobilities of both ion charges. The air's conductivity can be observed by measuring the current between the plates of a capacitor through which air is pulled. This is the basic idea behind air ion meters, and the best such devices can determine the concentration of each ion charge separately, and also determine the concentration of ions as a function of their mobility.[43] To build a simple air ion meter for classroom demonstrations, see Ref. 44.

## ACKNOWLEDGEMENTS

The author is grateful to the University of Alaska Anchorage for its support, and to the reviewers for their many suggestions.

## AUTHOR DECLARATIONS

The author has no conflicts to disclose.